\documentclass[apj]{emulateapj}

\def\ks{km~s$^{-1}$}

\def\mjup{M$_{\rm Jup}$}
\def\rjup{R$_{\rm Jup}$}

\def\msun{M$_{\odot}$}
\def\rsun{R$_{\odot}$}

\def\feh{[Fe/H]}

\def\vsini{$V_{\rm rot}\sin{i}$}

\def\lhs{LHS\,6343}

\def\kep{{\it Kepler}}
\def\dup{$6.5 \times 10^{-5}$} 
\def\k{9.6}
\def\ke{0.3}
\def\gam{-46.0}
\def\game{0.2}
\def\om{-23}
\def\ome{56}
\def\ecc{0.056}
\def\ecce{0.032}
\def\ma{0.370}
\def\mae{0.009}
\def\mb{0.30}
\def\mbe{0.01}
\def\d{36.6}
\def\de{1.1}
\def\fe{0.04}
\def\fee{0.08}
\def\ar{45.3}
\def\are{0.6}
\def\rr{0.226}
\def\rre{0.003}

\def\inc{89.50}
\def\ince{0.05}
\def\p{12.71382}
\def\pe{0.00004}

\def\u1k{0.190}
\def\u1ke{0.050}
\def\u1z{0.250}
\def\u1ze{0.090}
\def\u2k{0.2704}
\def\u2z{0.2548}
\def\mc{62.9}
\def\mce{2.3}
\def\ra{0.378}
\def\rae{0.008}
\def\arel{0.0804}
\def\arele{0.0006}
\def\imp{0.40}
\def\impe{0.04}
\def\va{13.88}
\def\vae{0.03}
\def\vb{14.63}
\def\vbe{0.06}
\def\bva{1.57}
\def\bvae{0.07}
\def\bvb{1.60}
\def\bvbe{0.07}

\def\dkp{0.74}
\def\dkpe{0.10}
\def\dz{0.5}
\def\dze{0.1}
\def\rrc{0.226}
\def\rrce{0.003}
\def\rc{0.833}
\def\rce{0.021}
\def\rrcs{0.051}
\def\rrcse{0.001}
\def\ja{10.10}
\def\jae{0.04}
\def\jb{10.59}
\def\jbe{0.06}
\def\ha{9.51}
\def\hae{0.04}
\def\hb{9.99}
\def\hbe{0.07}
\def\ksa{9.25}
\def\ksae{0.05}
\def\kb{9.70}
\def\kbe{0.08}
\def\gc{5.35}
\def\gce{0.02}
\def\rhoc{109}
\def\rhoce{8}
\def\rhos{6.6}
\def\rhose{0.4}
\def\ga{4.851}
\def\gae{0.008}

\begin{document}
\title{LHS\,6343\,C: A Transiting Field Brown Dwarf Discovered by the
  \emph{Kepler} Mission$^1$}  

\author{
John Asher Johnson\altaffilmark{2,3},
Kevin Apps\altaffilmark{4},
J. Zachary Gazak\altaffilmark{5},
Justin Crepp\altaffilmark{2},
Ian J. Crossfield\altaffilmark{6},
Andrew W. Howard\altaffilmark{7},
Geoff W. Marcy\altaffilmark{7},
Timothy D. Morton\altaffilmark{2},
Carly Chubak\altaffilmark{7},
Howard Isaacson\altaffilmark{7}
}

\email{johnjohn@astro.caltech.edu}

\altaffiltext{1}{ Based on observations obtained at the
W.M. Keck Observatory, which is operated jointly by the
University of California and the California Institute of
Technology. Keck time has been granted by Caltech,
the University of California and NASA.}
\altaffiltext{2}{Department of Astrophysics,
  California Institute of Technology, MC 249-17, Pasadena, CA 91125}
\altaffiltext{3}{NASA Exoplanet Science Institute (NExScI)}
\altaffiltext{4}{Cheyne Walk Observatory, 75B Cheyne Walk, Horley,
  Surrey, RH6 7LR, United Kingdom}
\altaffiltext{5}{Institute for Astronomy, University of Hawai'i, 2680
  Woodlawn Drive, Honolulu, HI 96822} 
\altaffiltext{6}{Department of Physics and Astronomy, University of
  California Los Angeles, Los Angeles, CA 90095}
\altaffiltext{7}{Department of Astronomy, University of California,
Mail Code 3411, Berkeley, CA 94720}

\begin{abstract}
We report the discovery of a brown
dwarf that transits one member of the M+M binary system \lhs\,AB
every 12.71 days. The transits were discovered using photometric data
from the {\it Kelper} public data release. The LHS\,6343 stellar system was
previously identified as a single high-proper-motion M dwarf. We use
adaptive optics imaging to resolve the system into two low-mass stars
with masses \ma$\pm$\mae~\msun\ and \mb$\pm$\mbe~\msun, respectively,
and a projected 
separation of $0\farcs55$. High-resolution spectroscopy shows that the more
massive component undergoes Doppler variations consistent with
Keplerian motion, with a period equal to the transit
period and an amplitude consistent with a companion mass of $M_C =
\mc \pm \mce$~\mjup. Based on our analysis of the transit light curve
we estimate the radius of the companion to be $R_C = \rc \pm
\rce$~\rjup, which is consistent with theoretical predictions of the
radius of a $> 1$~Gyr brown dwarf.
\end{abstract}


\section{Introduction}

Situated on the mass continuum between planets and hydrogen-burning
stars  are objects commonly known as brown dwarfs, with masses
spanning approximately 13\mjup\ up to
80~\mjup (assuming Solar metallicity). 
Since the first discoveries of these substellar objects
fifteen years ago \citep{oppenheimer95,basri96}, various surveys have 
found additional examples in  
numbers exceeding the population of known exoplanets\footnote{{\tt 
    http://spider.ipac.caltech.edu/staff/davy/ARCHIVE/}}. However, 
despite the large sample of brown dwarfs, very little is 
known about the physical properties or formation mechanism(s) of
these substellar objects \citep[e.g.][]{basri06,burgasser07,liu08b}. 

Most known brown dwarfs have been discovered as solitary objects by
wide-field, near-infrared (NIR) imaging
surveys\citep[e.g.][]{martin97,burgasser99,ukidss,delorme08}. Their
identification is often based on spectral typing, with physical
parameters derived from comparing photometric measurements to
substellar evolutionary models. Knowledge beyond spectral 
typing is limited by the difficulty in 
modeling the complex molecular features that dominate the spectra 
of cool dwarfs \citep[e.g.][]{allard01,cushing06}, a
problem that persists above the 
hydrogen-burning mass limit for M-type dwarfs \citep{maness07,johnson09b}. 

The preferred method of measuring physical
properties of substellar objects such as masses,
compositions and ages, is to study examples that are physically
associated with brighter main-sequence stars.  
By assuming both the brown dwarf and its host star formed at the same
time from the same molecular cloud, ages and chemical composition of
the companion can be tied to the properties of the brighter, more
easily characterized component \citep{liu05,bowler09}. However, brown
dwarfs in these favorable ``benchmark'' 
configurations are found in numbers far below the sample of
exoplanets, despite their relative ease of detection compared to
planet-mass companions. This observed feature
of the substellar mass distribution of bound companions is known as
the ``brown dwarf desert,'' and the barren region extends
over a wide swath around stars, extending from $\approx0.05$~AU out to
hundreds of AU \citep{marcy00,mccarthy04,grether06,johnson09rev}.

The existence of a deep minimum in the mass continuum between stars and
planets suggests that distinct formation mechanisms
operate at either mass extreme, one for stellar objects and one
for planets. However, the scarcity of objects
in the brown dwarf desert makes it difficult to determine where this
line should be 
drawn. For example, it is  unclear whether a 20~\mjup\ object in orbit
around a main-sequence star formed like a massive planet, or instead
should be considered part of the extreme low-mass tail of the stellar
initial mass function \citep{kratter10}. Furthermore, an issue as
fundamental as the mass-radius relationship below the hydrogen-burning
limit is largely unconstrained by observations. Understanding the
nature and origins of brown dwarfs requires a much larger sample of
detections. 

One promising avenue for increasing the sample of well-characterized
substellar companions is through wide-field photometric transit
surveys. Since the radii of objects are roughly constant from
1--100~\mjup\ \citep{baraffe98}, transit surveys are uniformly
sensitive to companions throughout the entire brown dwarf 
desert. The transit light curve, together with precise radial
velocities (RV), provide both the absolute mass (as opposed to minimum
mass, $M\sin{i}$) and radius of the companion, thereby directly testing
the predictions of interior structure models
\citep[][]{burrows97,torres08}. Once transits have been 
discovered, the door is opened up a wealth of follow-up
opportunities that can measure properties such as the brown dwarf's
albedo, temperature distribution, emission spectrum and atmospheric
composition \citep[see, e.g. the review by][]{charbonneau05}. Further, 
studying the  
distribution of physical characteristics of companions, and their
relationships to the characteristics of their host stars, can inform
theories of the 
origins of brown dwarfs in the same way that the statistical
properties of exoplanets inform theories of planet formation
\citep{fischer05b,torres08,johnson10c}. 

In this contribution we present the discovery and characterization of
a transiting brown dwarf orbiting a nearby low-mass star in the
\kep\ field. 

\section{Photometric Observations}

\subsection{\emph{Kepler} Photometry}

The {\emph Kepler} space telescope is conducting a continuous
photometric monitoring campaign of a target field near the
constellations Cygnus and Lyra. A 0.95-m aperture Schmidt telescope
feeds a mosaic CCD photometer with a $10^\circ \times 10^\circ$ field
of view \citep{koch10,kepler10}. Data reduction and analysis is
described in \citet{jenkins10a} and \citet{jenkins10b}, and the
photometric and astrometric data were made
publicly available as part of the first-quarter (Q0-Q1) data release. The Q0
data have a time baseline UT 2009 May 5-11, 
and the Q1 data span UT 2009 May 13 through 2009 June 16. 

Among the 156,000 Long Cadence stellar targets in the \emph{Kepler}
field is the nearby, high-proper-motion M dwarf  \lhs\ 
\citep[$\alpha = 19^h 10^m 14.31^s, \delta= +46^\circ 57^m
  25.0^s$;][]{reid04}. The 
photometric properties as listed in the Kepler Input Catalog
\citep[KIC;][]{kic} are given in Table~\ref{tab:photpars}. The
Q0-Q1 photometric time series of \lhs\  
contains a total of 2115 brightness measurements with a 29.4 minute
cadence and a median internal measurement precision of $7\times10^{-5}$
\citep{jenkins10b}. 

As part of a study of the photometric variability
of the closest stars in the \emph{Kepler} field, one of us
(K.A.) noted that the light curve of \lhs\ exhibits 
four deep, periodic dimming events spaced by 12.71~days. The light
curve depths are 
constant to 0.3~mmag and exhibit no obvious additional dimming at
intermediate 
periods, consistent with the signal of a transiting planet-sized
object. The
astrometry shows no shift in the center of light greater than 1
millipixel (4~mas, \citet{jenkins10c}), and there are no  
secondary eclipses evident at 
intermediate phases. It is therefore unlikely that the source of the 
dimming events is a background eclipsing binary (EB).

Examination of the Palomar Observatory Sky Survey (POSS) images shows no other stars at the
current position of \lhs, further ruling false-positives involving
an EB. The closest star in the archival
images is $7\farcs0$ to the West of \lhs, near the edge of the
\emph{Kepler} photometric aperture. However, the star is only 4\% as
bright as \lhs, meaning that if it is an eclipsing binary it would
have to nearly disappear to replicate the observed transit signal. 
This situation is ruled out by the lack of a large photocenter shift 
seen in the astrometric measurements. The \kep\ photometric
measurements phased at the 12.71 day period are shown in
Fig.~\ref{fig:lc}. 

\begin{deluxetable}{lcl}
\tablecaption{Observed Properties of LHS\,6343
\label{tab:photpars}}
\tablewidth{0pt}
\tablehead{
\colhead{Parameter} & \colhead{Value} & \colhead{Source} 
}
\startdata
$\alpha$   & 19\,10\,14.33 & KIC \\
$\delta$   & +46\,57\,25.5 & KIC \\
$\mu_\alpha$ (mas yr$^{-1}$) & -145 & KIC \\
$\mu_\delta$ (mas yr$^{-1}$) & -401 & KIC \\
$g^\prime$ & $14.03 \pm 0.02$ & KIC \\
$r^\prime$ & $13.06 \pm 0.02$ & KIC \\
$i^\prime$ & $12.07 \pm 0.02$ & KIC \\
$B_{\rm tot}$ & $15.009 \pm 0.025$ & \\
$V_{\rm tot}$ & $13.435 \pm 0.018$ &  \\
$K_{P,{\rm tot}}$ & $13.104 \pm 0.04$ & KIC \\
$J_{\rm tot}$        & $9.570 \pm 0.021$ & 2MASS \\
$H_{\rm tot}$        & $8.972 \pm 0.027$ & 2MASS \\
$K_{S,\rm tot}$      & $8.695 \pm 0.011$ & 2MASS \\
$\Delta J$   & $0.49 \pm 0.05$ & PHARO \\
$\Delta H$   & $0.48 \pm 0.05$ & PHARO \\
$\Delta K_S$ & $0.45 \pm 0.06$ & PHARO \\
& & \\
$J_A$     &  $\ja \pm \jae$ & \\ 
$J_B$     &  $\jb \pm \jbe$ & \\ 
$H_A$     &  $\ha \pm \hae$ & \\ 
$H_B$     &  $\hb \pm \hbe$ & \\ 
$K_{S,A}$     &  $\ksa \pm \ksae$ & \\
$K_{S,B}$     &  $\kb \pm \kbe$ &  
\enddata
\end{deluxetable}

\subsection{Nickel $Z$-band Photometry}

We observed the transit event predicted to occur on UT 2010 June 29
using the 1-m Nickel telescope at Lick Observatory on Mt. Hamilton,
California. We used the Nickel Direct Imaging Camera, which uses a
thinned Loral $2048^2$-pixel CCD with a $6.3\arcmin$ square field 
of view \citep{johnson08b}. We observed through a Gunn $Z$ filter,
used $2\times2$ 
binning for an effective pixel scale of $0\farcs37$~pixel$^{-1}$, and
a constant exposure time of 75 seconds. We used the slow readout mode,
with 34~s between exposures to read the full frame and reset
the detector. The conditions were clear with $\sim1\farcs0$
seeing. We began observing as soon as possible after sunset at
an airmass of 1.4 and observed continuously for 5.4 hours bracketing
the predicted transit midpoint, ending at an airmass of 1.05. 

\begin{figure}[!t]
\epsscale{1.1}
\plotone{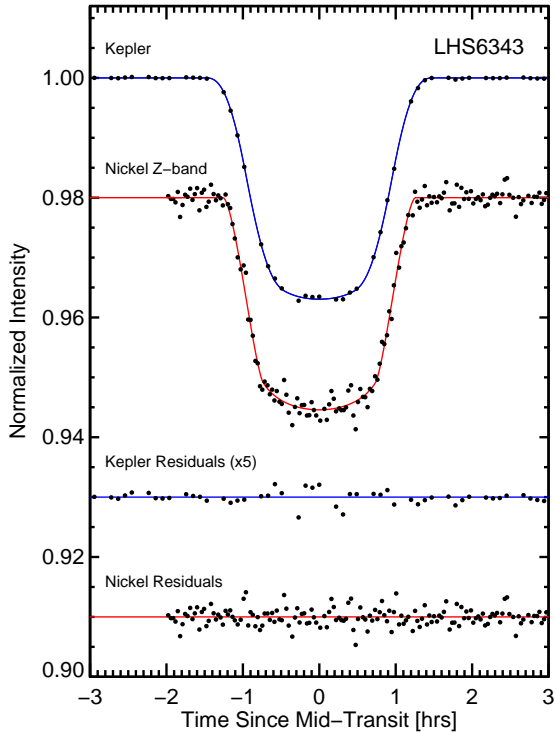}
\caption{The \kep\ (upper) and Nickel (blue) light curves, phased at
  the photometric period. The best-fitting light curve models are
  shown for each data set (see \S~\ref{sec:lcfit}), and the residuals
  are shown beneath each 
  light curve. The Kepler residuals have been multiplied $\times$5
  for clarity.  \label{fig:lc}}  
\end{figure}

We measured the instrumental magnitude
of \lhs\ with respect to the four brightest comparison stars in the
field using an aperture width of 23 pixels and a sky annulus with an
inner and outer radius of 28 and 33 pixels, respectively.  We
converted the Nickel timestamps to BJD$_{UTC}$ using the techniques of 
\cite{eastman10} to be consistent with the \kep\ data. The
Nickel photometric measurements phased at the 12.71 day period are
shown in Fig.~\ref{fig:lc}.

\section{Radial Velocities and Orbit Solution}
\label{sec:rv}

We obtained spectroscopic observations of \lhs\ at Keck
Observatory using the HIRES spectrometer with a resolution of $R
\approx 55,000$, with the standard iodine-cell setup used by the
California Planet Survey \citep{howard10a}. The transit depth, together
with a rough stellar 
radius estimate of 0.4~\rsun\ appeared consistent with a planet
with a radius of $\sim0.7$~\rjup. In anticipation of a low-amplitude
Doppler signal we initially used 45 minute exposures through the
iodine cell and C2 decker for sky-subtraction. The resulting
signal-to-noise ratio (S/N) was $\approx90$ at 5500~\AA, near the
center of the iodine absorption region. 

A cross-correlation analysis of the first two observations revealed
two peaks separated by $\sim10$~\ks. The lack of a coincident
background star in the POSS images suggests that the second set of
lines must be from a physically-associated binary companion and that
LHS\,6343 is a double-lined spectroscopic binary. Adaptive optics
observations described in \S~\ref{sec:ao} confirmed the existence of a
wide binary companion at a projected separation of
$0\farcs55$. Hereafter, we refer to the more massive component as Star
A, and the less massive component as Star B.

In order to discern which component of the binary system is
transited by a companion we made subsequent HIRES observations 
with a position 
angle oriented along the binary axis to ensure the light from both
stars fell within the slit. The cross-correlation analysis of the
third observed spectrum revealed that the deeper set of absorption lines 
shifted by $\sim10$~\ks\ with respect to the first observation,
indicating that Star A is orbited by a massive companion.  

Our remaining HIRES spectra were obtained without the iodine cell
and with 3-minute exposure times. We measured the radial velocity of
Star A using the cross-correlation analysis described by 
\citet{johnson04b}. However, we modified the procedure by 
constructing a double-lined cross-correlation
template. We began with an iodine-free spectrum of the HD\,265866
(M3V) for star A, and added to this template a scaled, shifted version of
itself to represent the spectrum of star B. For each observation we
adjusted the scaling and Doppler-shift of spectrum B with respect to
spectrum A until the cross-correlation peak was maximized. We then
measured the centroid of the optimized cross-correlation function by
fitting a parabola to the region near the resulting single peak. 

\begin{figure}[!t]
\epsscale{1.2}
\plotone{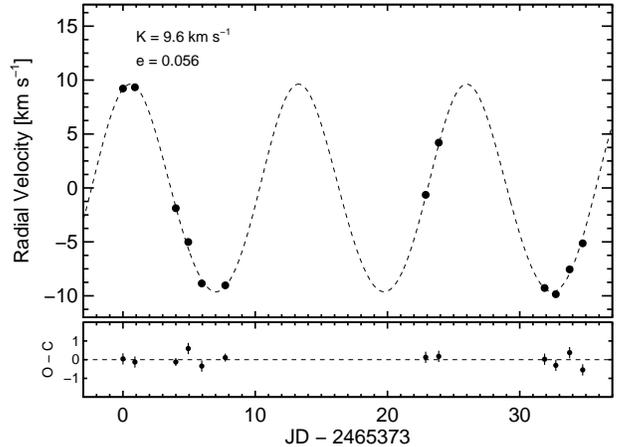}
\caption{Keck/HIRES radial velocity measurements of \lhs\,A. The
best-fitting Keplerian orbit solution is shown as a dashed line. The 
  systemic velocity, $\gamma = \gam \pm \game$~\ks, has been
  subtracted for clarity. The lower panel shows the RV residuals
  about the best-fitting orbit. \label{fig:rv}} 
\end{figure}

We corrected for shifts in the HIRES detector by using the telluric
lines in the 630~nm $\alpha$-band as a wavelength reference. 
We measured the absolute radial velocities of \lhs\,A with respect to
the absolute radial
velocity of HD\,265866 (M3V; $V_r = +22.97 \pm 0.50$~\ks; Chuback et
al., in prep), and the resulting
RV is corrected to the Solar System barycenter using the velocity
corrections computed by the CPS data reduction pipeline.  The full
time series of radial velocity measurements is displayed in
Fig.~\ref{fig:rv}. The radial velocities are also listed in
Table~\ref{tab:rv}, 
along with the heliocentric Julian Dates of observation and internal
measurement errors.

We searched for the best-fitting Keplerian orbit solution using the
partially linearized, least-squares fitting procedure described in
\citet{wrighthoward} and implemented in the IDL software package {\tt
RVLIN}\footnote{http://exoplanets.org/code/}. We fixed the period and
mid-transit time based on the light-curve analysis in
\S~\ref{sec:lcfit}, which leaves only four free parameters: the
velocity semiamplitude ($K_A$), argument of periastron ($\omega$),
systemic velocity ($\gamma$) and eccentricity. We find that the RVs
are described well by a nearly 
circular orbit ($e=\ecc \pm \ecce$) with a velocity semiamplitude $K = \k
\pm \ke$~\ks. The full spectroscopic orbit is given in
Table~\ref{tab:starpars} and shown in Fig.~\ref{fig:rv}. 

The parameter uncertainties were estimated using
a bootstrap Monte Carlo algorithm. For each of 5000 realizations of
the data, the measured RVs are perturbed by adding residuals
randomly drawn from about the best-fitting orbital solution, with
replacement. We chose this technique over an MCMC analysis out of
concern that our RV measurement uncertainty is dominated by systematic
errors related to imperfect treatment of the second set of absorption
lines, rather than photon noise. Thus, instead of assuming the RVs are
normally distributed about the model, we use the residuals themselves
as an estimate of the noise model.

\section{Palomar Adaptive Optics Imaging}
\label{sec:ao}

We acquired near-infrared images of LHS 6343 on 
UT 2010 June 29 using the Palomar 200-inch telescope adaptive optics
system and PHARO camera \citep{hayward01, troy00}. These 
diffraction-limited observations spatially resolve the target into a 
binary, as shown in Fig. 3.  We used these observations to calculate 
the relative brightness of each component of the visual binary and to 
subsequently constrain the range of possible masses 
(see \S~\ref{sec:starpars}). Our results for
differential magnitudes 
in each of the $J$,$H$,$K_S$ filters are listed in
Table~\ref{tab:photpars}. 

We used aperture photometry To measure the differential magnitudes of
the two stars. However, given
their relatively small angular separation, special care was taken to
account for cross-contamination between the components. To remove the
majority of flux contributed by the neighboring star, we used the
spatial symmetry in the images. The amount of contaminating starlight
was estimated by summing the counts over a region the same size as the
photometric aperture located on the side opposite the star of
interest. 

\begin{figure}[!t]
\epsscale{1.1}
\plotone{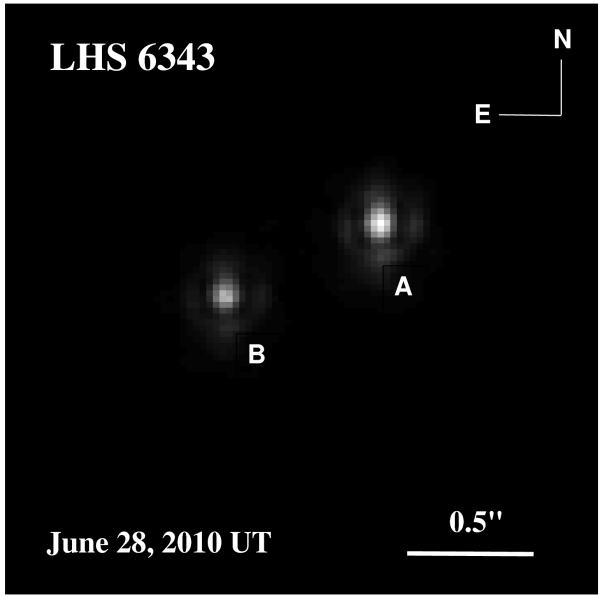}
\caption{PHARO $K_S$-band adaptive optics image of LHS\,6343, showing the
  two M-type components of the system. Star A is to the upper right of
  the image and  Star B is to the  lower left. \label{fig:ao}} 
\end{figure}

Once the contaminating light from the neighbor star is removed, we
find that the photometric precision is limited to several percent by
uncertainties resulting from the subtraction residuals, as well as CCD
non-linearity and PSF centroiding. These errors sources contribute
similarly to the overall uncertainty and were added in quadrature,
neglecting any correlations between PSF centroiding errors and
contamination removal, 
which we found to be comparatively small. To this, we also added in
quadrature the standard deviation in the mean flux ratio of the
companions over the 20 images acquired in each bandpass. Observations
in the $K_S$ band have a slightly larger uncertainty than in J and H,
since the binary separation subtends a smaller angle on the sky in
units of resolution elements. 

\section{Stellar Properties}
\label{sec:starpars}

\begin{deluxetable}{lll}
\tablecaption{Radial Velocities for LHS\,6343\label{tab:rv}}
\tablewidth{0pt}
\tablehead{
\colhead{JD} &
\colhead{RV} &
\colhead{Uncertainty} \\
\colhead{-2440000} &
\colhead{(km~s$^{-1}$)} &
\colhead{(km~s$^{-1}$)} 
}
\startdata
15373.095 &  -38.40 &  0.21 \\
15373.998 &  -37.51 &  0.17 \\
15377.078 &  -49.42 &  0.50 \\
15377.099 &  -49.56 &  0.45 \\
15378.030 &  -52.56 &  0.41 \\
15379.052 &  -56.39 &  0.44 \\
15380.827 &  -55.83 &  0.48 \\
15380.831 &  -54.84 &  0.47 \\
15395.983 &  -47.92 &  0.57 \\
15396.970 &  -42.63 &  0.51 \\
15404.974 &  -55.90 &  0.59 \\
15405.821 &  -56.32 &  0.46 \\
15406.865 &  -53.69 &  0.51 \\
15407.854 &  -51.31 &  0.57
\\
\enddata
\end{deluxetable}

The physical properties of the two stellar components of the wide binary,
hereafter Star A and Star B, are of central importance to measuring
the properties of the substellar companion, \lhs\,C. The mass of the
companion is related to the mass of Star A ($M_A$) and the companion's
radius depends on the stellar radius, $R_A$. The luminosity of 
Star B is also important for the measurement of the companion's
radius, as its contribution to the total flux of the system dilutes
the transit depth. In fact, for the specific case of \lhs\ the
precision with which we can measure the companion radius will depend
critically on our estimate of the ``third light'' contribution of Star
B, rather than the photometric precision of the transit light curve 
\citep{irwin10}.

Because low-mass stars spend nearly their entire lives close to the
zero-age main sequence, their observed properties are a function of
two physical characteristics: mass and, to a lesser extent,
metallicity. There are two widely-used methods estimating the masses
of M dwarfs. The first involves a comparison of observed properties
such as absolute magnitude and color index, or luminosity and
effective temperature, to tabulated stellar
evolution models, such as those calculated by
\citet{baraffe98}. However, studies of low-mass eclipsing binaries
(EB) have demonstrated that these models systematically under-predict
stellar radii \citep{ribas06, lopez07}, even in cases when stellar
activity should play a minimal role in shaping stellar structure
\citep{torres07}.   

The other method makes use of empirical relationship between
the near-infrared luminosity of a star and its mass, as parametrized
by \citet{delfosse00}. We use the empirical
relationships almost exclusively in order to avoid any
systematic errors in the stellar evolution models. However, since the
mass-luminosity relationships 
require absolute magnitudes, the distance to the star must be known
to accurately estimate the stellar mass. Unfortunately,
there is no published trigonometric parallax for \lhs, and the
spectroscopic parallax of \citet{reid04} is unreliable because it is
based on the total magnitude and colors of the binary system, rather
than the individual stars.

While the binarity of LHS\,6343 in some regards poses a challenge,
having two stars 
with the same age and chemical composition, together with the available
photometric measurements, provide a unique opportunity to determine the
physical characteristics of the two components. As we will
demonstrate, the luminosity difference between the two stars
constrains the mass ratio, while the total luminosity constrains the
total mass, distance and metallicity of the system. 

An additional constraint on the mass is provided by the shape of the
transit light curve. The slope of the ingress/egress yields the scaled
semimajor axis $a/R_A \equiv a_R$, which is related to the density of 
Star A through Kepler's third law
\citep[e.g.][]{seager03,soz07,winn08r}. However, the true value of $a_R$
depends on the true transit depth $(R_C/R_A)^2$, which in turn is
related to the amount of dilution in the light curve due to Star B. 

In what follows we first relate the masses of Star A and Star B to the
available observables. We then estimate the flux contribution of Star
B in the \kep\ bandpass, which provides a refined estimate of the
stellar masses. The iteration of this procedure yields the stellar
parameters of Star A, which allows us to estimate the physical
characteristics of the transiting object \lhs\,C.

\begin{deluxetable}{rrrrrr}
\tablecaption{Polynomial Coefficients for Magnitude-Mass and
  Mass-Radius Relationships 
\label{tab:Mcoefs}}
\tablewidth{0pt}
\tablehead{
\colhead{$j$} &
\colhead{$b_J$} & 
\colhead{$b_H$} & 
\colhead{$b_{K_S}$} & 
\colhead{$b_V$}  & 
\colhead{$b_R$}
}
\startdata
0 & 14.888   & 13.211  & 13.454  & -9.229  & 0.000  \\
1 & -74.375  & -57.464 & -67.439 & 5.017   & 1.268  \\
2 & 376.72   & 271.11  & 338.43  & -0.6609 & -1.013 \\
3 & -1089.6  & -762.12 & -976.27 & 0.03314 & 0.9391 \\
4 & 1601.6   & 1103.9  & 1433.9  & ...     & ...    \\
5 & -935.37  & -640.80 & -838.03 & ...     & ...    \\
\enddata
\end{deluxetable}

\subsection{Stellar Masses}
\label{sec:masses}
The most useful data available to us comprises seven photometric
measurements: the total near-infrared 
magnitudes in the 2MASS catalog denoted by ${\mathcal{T}_i}$
where $i = J,H,K_S$ corresponds to the three bands, respectively;
the magnitude differences ${\Delta J, \Delta H, \Delta K_S}$ from our
AO imaging, denoted by $\Delta_i$. We also have the total Johnson $V$-
and $B$-band magnitudes $V_{tot} = 13.435 \pm 0.018$ and $B_{tot}
=15.009 \pm 0.025$. Our $V_{tot}$ agrees well with 
the value listed in the TASS catalog, $V_{\rm tot} = 13.38 \pm 0.24$ 
\citep{tass}. These magnitudes yield the system color $(B-V)_{\rm tot}
= 1.574 \pm 0.031$.  

The individual apparent magnitudes in the $i$th NIR band (JHK) of Star
A ($m_{i,A}$) and Star B ($m_{i,B}$) are related to the total
magnitudes $\mathcal{T}_i$ and magnitude differences $\Delta_i$ through

\begin{eqnarray}
m_{i,A} &=& 2.5\log_{10}(1 + 10^{0.4 \Delta_i}) + \mathcal{T}_i \nonumber \\
m_{i,B} &=& \Delta_i + m_{i,A}
\label{eqn:appmag}
\end{eqnarray}

\noindent Eqns~\ref{eqn:appmag} give three NIR apparent
magnitudes for each star. These six NIR photometric measurements can
be related to the stellar  masses ($M_A, M_B$) through the following
equations 

\begin{eqnarray}
\label{eqn:modelmag}
m_{i}(M,d) &=& \mathcal{M}_i(M) + 5(\log_{10}d - 1)
\end{eqnarray}

\noindent The functions ${\mathcal{M}_i(M)}$ give the absolute
magnitudes in the NIR bands as a function of stellar mass, $M$, as
determined by inversion of the empirical relationships of
\citet{delfosse00}, which we approximate with the polynomial 

\begin{equation}
\mathcal{M}_i(M) = \sum_{j=0}^{5} b_{i,j} M^j
\end{equation}

\noindent The coefficients $\{b_i\}$ are listed in
Table~\ref{tab:Mcoefs} for the $i = \{J,H,K_S\}$ bands. 

Under the assumption that the binary components share the same chemical
composition, they  should reside at the same distance from the average
main sequence in the $\{V-K_S$, $M_{K_S}\}$ plane
\citep[][hereafter JA09]{johnson09b}. JA09 provide a relationship
between a star's metallicity, [Fe/H]~$\equiv F$, and its ``height''
above the Solar-neighborhood mean main sequence, $\Delta M_K$. Since
the stars share the same composition the must lie on the same
isometallicity contour: their 
$V-K_S$ colors must be consistent with the same value of $\Delta
M_K$, while the individual $V$-band luminosities must be consistent
with the measured $V_{\rm tot}$. This constraint can be expressed as

\begin{eqnarray}
V_{\rm tot}(M_A, M_B, d, F) &=& \mathcal{M}_{\rm V}(M_A, F) \nonumber \\ 
&-& 2.5\log_{10}(1 +
10^{0.4[\mathcal{M}_{\rm V}(M_A,F)-\mathcal{M}_{\rm V}(M_B,F)]}) \nonumber \\ 
&+&  5\log_{10}d - 5,
\label{eqn:vtot}
\end{eqnarray}

\noindent where the absolute
V-band magnitude,
$\mathcal{M}_{\rm V}(M,F)$, is related to stellar mass $M$ and
metallicity $F$ by 
inverting the photometric metallicity calibration of
JA09\footnote{We use the JA09
  relationship rather than the \citet[][SL10]{sl10} calibration
  because the former provides a better match to the mean metallicity
  of the Solar neighborhood. In principle, the SL10
  relationship would serve our purposes just as well since both
  $V$-band metallicity relationships provide a means of relating
  $V_{\rm tot}$ to the stellar masses under the constraint that both stars 
lie on the same isometallicity contour in the $\{V-K_S$, $M_{K_S}\}$
plane. The only difference is the exact value of 
[Fe/H], which will be $\approx 0.1$ dex lower using the
SL10 calibration.}. We approximate this inversion using the
polynomial 

\begin{equation}
\mathcal{M}_{\rm V}(M,F) = \sum_{j=0}^{3} b_{\rm{V},j} \left[\mathcal{M}_{K_S}(M) + \left(\frac{F - 0.05}{0.55}\right)\right]^j 
\label{eqn:fe}
\end{equation}

\noindent The coefficients $\{b_V\}$ are listed in Table~\ref{tab:Mcoefs}.

In addition to the apparent magnitudes, the transit light curve
provides 
an additional constraint on $M_A$ through the
scaled semimajor axis $a/R_A \equiv a_R$. This quantity is related to
the mass and radius of Star A, and the period and mass of \lhs\,C,
through  

\begin{eqnarray}
a_R(M_A, M_C, P) &=&
\left(\frac{G}{4\pi^2}\right)^{1/3} \frac{M_A^{1/3}}{R_A(M_A)}
P^{2/3} \left(1 + \frac{M_C}{M_A}\right)^{1/3}, \nonumber \\
\label{eqn:ar}
\end{eqnarray}

\noindent where $R_A(M)$ is a function relating the stellar
radius and mass. We use a polynomial fit to the masses and radii of
well-characterized ($\{\sigma_M,\sigma_R\} < 3$\%) low-mass eclipsing
binaries tabulated by \citet{ribas06}

\begin{equation}
R_A(M) = \sum_{j=0}^{3} b_{R,j} M^j
\label{eqn:ribas}
\end{equation}

\noindent The coefficients $\{b_R\}$ are listed in Table~\ref{tab:Mcoefs}. 

The optimal stellar parameters can be obtained by minimizing the
fitting statistic

\begin{eqnarray}
\chi^2_{\rm tot} &=&
\sum_{i=1}^3
\left(
\frac{m_{i,A} - m_i(M_A,d)}{\sigma_{m_{i,A}}}\right)^2 \nonumber \\
&+&
\sum_{i=1}^3
\left(
\frac{m_{i,B} - m_i(M_B,d)}{\sigma_{m_{i,B}}}\right)^2 \nonumber \\
&+&
\left( \frac{V_{\rm tot} - V_{\rm
    tot}(M_A,M_B,d,F)}{\sigma_{V_{\rm tot}}}\right)^2 \nonumber
\\ 
&+&
\left( \frac{a_R - a_R(M_A, M_C, P)}{\sigma_{a_R}}\right)^2
\label{eqn:chi}
\end{eqnarray}

\noindent However, in order to obtain the best-fitting parameters we must
determine the corrected values of the transit depth and $a_R$ by
accounting for the flux contribution of Star B, as described in the
following section.

\subsection{Flux Contribution of Star B}
\label{sec:starB}

Since both stars fall within the {\it Kepler} and Nickel photometric
aperture, we must estimate the flux contribution
of Star B in both the $K_P$ and $Z$ bandpasses. The transformation
between the Johnson $B$ and $V$ magnitudes and the Kepler magnitude
$K_P$ is given 
in the Kepler Guest Observer web page\footnote{{\tt
    http://keplergo.arc.nasa.gov/CalibrationZeropoint.shtml}}. However,
while we were able to measure individual $V$ magnitudes for both
stars based on their total $V$ magnitude and derived physical
properties using Eqn.~\ref{eqn:fe}. However, there is no suitable
relationship between the available observables and the individual $B$
magnitudes. We are therefore forced rely on stellar model grids to
estimate the $B-V$ colors of stars A and B.

For this task, we selected the ``Basic Set'' of stellar model grids
from the Padova group for Solar composition and an age of 5~Gyr
\citep{girardi02},  
which give predictions for the Johnson $B$ and $V$ magnitudes as a
function of stellar mass. We selected the Padova models because the
\citet{baraffe98} model grids  do not give fluxes in the $B$ or $Z$  
bandpasses. We tested the reliability of the Padova model grids 
using a sample of 6 metal-rich stars from \citet{johnson09b}. We found 
that the models under-predict the $B-V$ colors of our calibration
stars by $0.20 \pm 0.07$, independent of stellar mass over 
a range of approximately 0.1~\msun to 0.6~\msun. We applied
this correction to the Padova models to obtain a refined relationship
between stellar mass and $B-V$ color, and the uncertainty of this
offset was folded into the final uncertainties in the colors.

Using the estimates of the $B-V$ colors of Star A and Star B, together
with the individual $V$-band magnitudes from Eqn.~\ref{eqn:fe} we can
estimate the relative magnitudes of the stars in the \kep\ bandpass,
$\Delta K_P$, as a function of the stellar masses. We then use this
value of $\Delta K_P$ to correct the \kep\ light curve parameters.
For the Nickel light curve, we use the SDSS $z$-band magnitude from
the Padova model as a proxy for the Gunn $Z$ filter. 

Our reliance on low-mass stellar model grids for the magnitude
differences is less than ideal, particularly in optical bandpasses, 
because of the difficulty in properly treating molecular opacities at
optical wavelengths \citep[e.g.][]{baraffe98}. In the analysis that
follows, we attempt to ameliorate this imperfect knowledge by using
large uncertainties for $\Delta K_P$ and $\Delta Z$, which are 
propagated throughout our analysis and reflected in the confidence
intervals for our derived system properties. We discuss the impact of
our model-based magnitude differences in \S~\ref{sec:impact}.

\subsection{Light Curve Analysis}
\label{sec:lcfit}

We fitted the \kep\ and Nickel light curves using the analytic eclipse
model of \citet{mandelagol} to compute the integrated flux from the
uneclipsed stellar surface as a function of the relative positions of
the star and planet. The parameters were the period $P$, inclination
$i$, ratio of the companion and stellar radii $R_C/R_A$, the scaled
semimajor axis $a_R$, time of mid-transit $T_{\rm mid}$, and the
parameters describing the limb-darkening of the star. 
For the limb-darkening we used the quadratic approximations of
\citet{sing10} and \citet{claret04} for the \kep\ and Z bands,
respectively. In our model we fixed the quadratic terms $u_2$ at the
tabulated values for each filter, and allowed the linear term $u_1$ to
vary under a penalty of the form exp$[-(x -\mu_x^2)/\sigma_x^2]$, which
  is added to the fitting statistic. This treatment of the limb darkening
was made based on the additional structure seen in the residuals of 
the \kep\ light curve fit. Allowing the limb-darkening parameters to float
freely results in unphysical values. 

To properly model the light curves we made two modifications to
the typical light curve analysis. First, we corrected for the
third-light component by adjusting the normalized flux level from the
Mandel \& Agol light curve model, $f_{mod}$, such that

\begin{equation}
f_{\rm corr} = \frac{f_{\rm mod} + 10^{-0.4 \Delta K_P}}{1+ 10^{-0.4 \Delta K_P}}
\label{eqn:fcorr}
\end{equation}

\noindent In our fitting procedure, we
treat $\Delta K_P$ as a free parameter under the
normally-distributed, prior constraint
$\mathcal{N}(\Delta K_P,\sigma_{K_P})$ based on the estimate of
the magnitude differences described in 
\S~\ref{sec:starB}. We use a similar prescription for the flux
contribution of Star B to the Nickel
$Z$ bandpass. Our other modification involved rebinning
the analytic light curve to match the 29.4-minute cadence of the data
using a method similar to that of \citet{kipping10}. 

In addition to modeling the companion transit, we also
fitted a slowly-varying function to the out-of-transit portion of
the Nickel light curve to account for differential extinction from
the Earth's atmosphere. To account for
time-correlated noise in the data we used a Daubechies fourth-order
wavelet decomposition likelihood  
function following the technique described by \citet{carter09}.
Wavelet decomposition 
provides increased confidence in the derived parameter uncertainties
over the traditional $\chi^2$ likelihood by allowing parameters that
measure photometric scatter (uncorrelated white noise $\sigma_w$, and
$1/f$ red noise $\sigma_r$) to evolve as free parameters.  The
method recovers the traditional $\chi^2$ fitting statistic in the
case of $\sigma_r = 0$, and when $\sigma_w$ is fixed at a value
equal to the characteristic measurement error.

We determined the best-fitting model parameters and their
uncertainties using a Markov Chain Monte Carlo (MCMC)
analysis with a Gibbs Sampler \citep{geman84,ford05,winn09c}. 
The MCMC fitting algorithm was implemented using the Transit Analysis
Package (TAP; Gazak \& Johnson 2011, in prep.), a graphical user
interface--driven analysis tool written  in the \emph{Interactive Data
  Language} 
(IDL). We constructed 10 chains containing $10^6$ links using initial
conditions based on a simple least-squares fit to the phased
photometry. We chose step sizes such that 30--40\% of the steps are
accepted. We discarded the initial $10^5$ links in each parameter
chain to allow for ``burn-in'' before combining the chains.  The
resulting chains of parameters form the posterior probability
distribution, from which we select the 15.9 and 84.1 percentile levels
in the cumulative distributions as the ``1$\sigma$'' confidence
limits. In most cases the posterior probability distributions were
approximately Gaussian, and we therefore report only symmetric error
bars for simplicity. The final, iterative fitting procedure and
derived parameters are presented in \S~\ref{sec:system}

\begin{deluxetable}{ccc}
\tablewidth{0pt}
\tablecaption{LHS 6343 Ephemeris}
\tablehead{
\colhead{T$_{\rm mid}$ (BJD - 2450000.0)}
& \colhead{T$_{\rm mid}$ - Ephemeris}
& \colhead{Telescope}
}
\startdata
4957.216636 $\pm$  0.000073 &    0.0000020 $\pm$ 0.00012 & K     \\
4969.930410 $\pm$  0.000099 &    -0.000045 $\pm$ 0.00014 & K      \\
4982.64437   $\pm$  0.00013     &     0.000093 $\pm$ 0.00016 & K    \\
4995.35807   $\pm$  0.00015   &     -0.000028 $\pm$ 0.00018 & K     \\
5376.77274   $\pm$  0.00025     &     -0.0000050 $\pm$ 0.00027 & N 
\enddata
\tablecomments{K -- \kep, N -- Nickel Z-band}
\label{tab:midtimes}
\end{deluxetable}

\section{Physical Properties of the \lhs\ Stellar System}
\label{sec:system}

We solve for the physical parameters of the \lhs\ system using the
following iterative procedure:

\begin{enumerate}
\item Fit the light curves to obtain the scaled semimajor axis,
  $a_R$. We initially use $\Delta K_p = 0.7$, $\sigma_{K_P} = 0.1$,
  $\Delta Z = 0.5$, and $\sigma_Z = 0.1$ in Eqn.~\ref{eqn:fcorr} and
refine these values with subsequent iterations.
 
\item Estimate $M_C$ based on our Keplerian fit to the RV time
  series by solving 

\begin{equation}
M_C^3\sin^3i = \frac{K^3 P }{2\pi G} (M_A+M_C)^2\sqrt{1-e^2}
\label{eqn:mc}
\end{equation}

  We initially assume $M_C \ll M_B$, and relax this
  assumption as $M_A$  is revised.

\item Use $a_R$ and $M_C$ to estimate $M_A$ and $M_B$ by minimizing
  Eqn.~\ref{eqn:chi}. 

\item Test for convergence in $M_A$, $M_B$, $M_C$ and $\Delta
  K_P$ by comparing the current values to those of the previous
  iteration. If the change is larger than 10\% of the parameter
  uncertainties, then go to step 1.  
\end{enumerate}

\begin{figure}[!t]
\epsscale{1.2}
\plotone{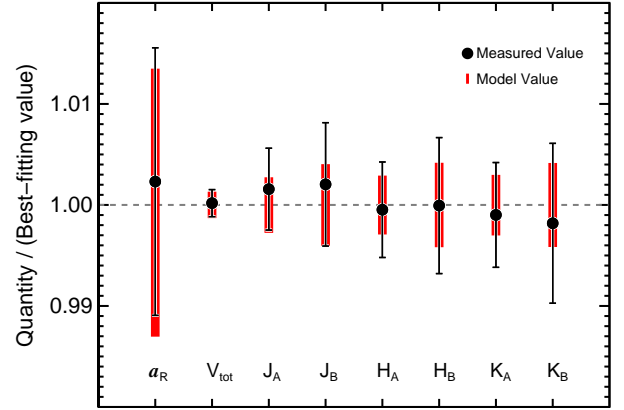}
\caption{Comparison of the observed stellar and transit properties
  (filled circles with error bars),
  and the corresponding quanities predicted by our model
  (red swaths; Eqns~\ref{eqn:modelmag}, \ref{eqn:vtot} and
  \ref{eqn:ar}). In order to show all parameters on the same scale, the
  quantities have been divided by the best-fitting model
  values. 
 \label{fig:bestfit}}       
\end{figure}

\begin{figure}[!t]
\epsscale{1.2}
\plotone{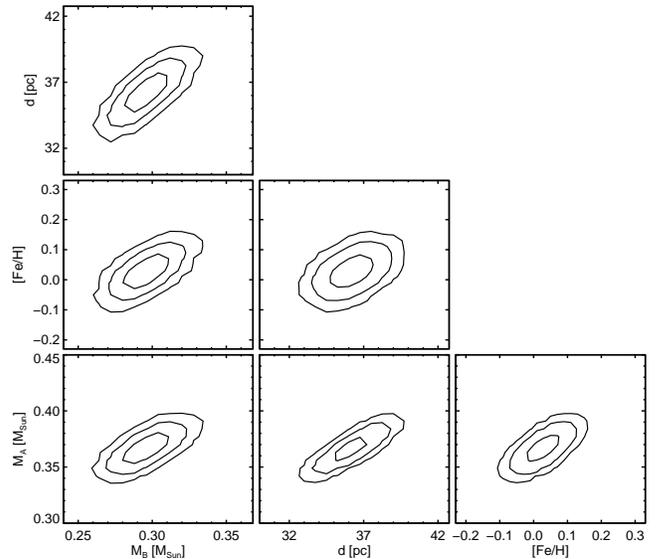}
\caption{Each panel shows the joint, posterior pdfs for two parameters
  at a time, marginalized   over the remaining parameters. The
  contours show the iso-probability 
  levels corresponding to $\{68.2, 95, 99.7\}$\%
  confidence. \label{fig:pdf}}  
\end{figure}
 
\noindent We find that convergence is rapid, requiring only three iterations.

Our best-fitting stellar parameters are $M_A = \ma \pm \mae $~\msun,
$M_A = \mb \pm \mbe $~\msun, $d =\d \pm \de$~pc and [Fe/H]~$=\fe \pm \fee$. 
The fit results in $\chi^2_{\rm tot} = 2.1$ with 8 data points and 4
free parameters, indicating that our fit is acceptable but that our 
photometric measurement uncertainties are likely overestimated. 
The best-fitting model values are compared to
the observed quanities in Fig.~\ref{fig:bestfit}. The
marginalized, posterior probability density functions (pdf) are shown
in Fig.~\ref{fig:pdf}.

Our analysis of the light curve results in $R_C/R_A = \rr \pm
\rre$ and $a_R = \ar \pm \are$, with $\Delta K_P = \dkp \pm
\dkpe$ and $\Delta Z = \dz \pm \dze$. The best-fitting transit model
is shown in Fig.~\ref{fig:lc} for both the \kep\ (blue) and Nickel
data (red), along with the residuals. For both data sets we recover 
$\sigma_r = 0$, consistent with no red noise contamination. However, we
do see additional structure in the in-transit residuals of the
\kep\ fit. We do not 
know the source of this increased scatter. Because there
are so few points, and because the data do not appear to be time-correlated 
during any single transit event, we found that the extra scatter is
accounted for by allowing the fitting procedure to increase the
white-noise component, $\sigma_w$. We find
$\sigma_w = 1.1 \times 10^{-3}$ (1.8-minute
cadence) for the Nickel 
data and $\sigma_w = 1.1 \times 10^{-4}$ (29.4-minute cadence) for the
\kep\ data. 

\subsection{The Radii of \lhs\,A and \lhs\,C}

Given the mass of Star A, we can estimate the stellar radius
by evaluating Eqn.~\ref{eqn:ribas} which results in a stellar radius
$R_A = \ra \pm \rae$~\rsun. 
The radius ratio from the light curve analysis, together with $R_A$
yields the radius of \lhs\,C, $R_C = \rc \pm \rce$~\rjup. The other
physical properties of the brown dwarf are listed in
Table~\ref{tab:starpars}. 

\subsection{The Dependence of $R_C$ on the Flux Contribution of \lhs\,B}
\label{sec:impact}

In \S~\ref{sec:starB} we describe our method of estimating the flux
contribution of Star B, as parametrized by the magnitude differences
$\Delta K_P$ and $\Delta 
Z$. These values are of central importance to the determination of the
corrected values of $a/R_A$ and $R_C/R_A$ as derived from the analysis
of the light curves, and hence the physical properties of the brown
dwarf. Unfortunately, we do not have spatially--resolved photometry of
the binary stars, so we were forced to rely on theoretical stellar
models to estimate the magnitude differences. Owing to incomplete
knowledge of molecular opacities, particularly in optical bandpasses,
the model grids provide only rough estimates of the true flux
contribution of Star B. To encapsulate this imperfect knowledge in our
analysis, we used relatively large uncertainties (0.1~mag) for $\Delta
K_P$ and $\Delta Z$. These errors are propagated throughout our MCMC
analysis and are  
reflected in the final physical properties of the \lhs\ system.

\begin{figure}[!t]
\epsscale{1.2}
\plotone{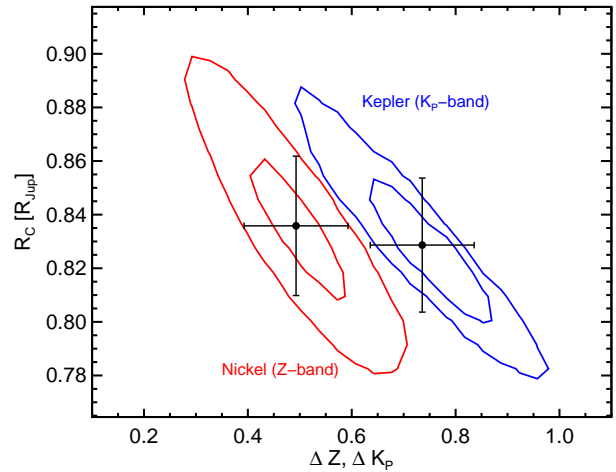}
\caption{The dependence of $R_C$ on $\Delta K_P$ (upper, blue) and
  $\Delta Z$ (lower, red), based on the marginalized posterior
  pdfs. The contours represent the 68.2\% and 95\% 
  confidence regions. The two filled circles with error bars show 
  the final values of $R_C$, $\Delta K_P$ and $\Delta Z$ from our
  analysis, and the error bars represent the corresponding 68.2\% 
  confidence bounds in each dimension. \label{fig:rp}}   
\end{figure}

As a test, we performed independent fits to the \kep\ and
$Z$-band light curves. We found that the transit properties from the
two analyses agreed extremely well, differing by only a fraction of a
$\sigma$ in each value. Figure~\ref{fig:rp} illustrates
the dependence of $R_C$ on $\Delta K_P$ and $\Delta Z$, and compares
the value of $R_C$ measured from the independent light curve fits.
Based on this figure we feel that our value of $\Delta Z$ is
reasonable since its value shouldn't be less than 
$\Delta J = 0.49$. Similarly, and $\Delta Z$ shouldn't exceed our
estimate of $\Delta K_P$, and $\Delta K_P$ shouldn't be much larger
than our measured $\Delta V = 0.74$. These results, along with the
close agreement between the two measurements of $R_C$, provide
additional confidence that our model-based magnitude 
differences are valid.  

\subsection{Searching for Transit Timing Variations}

To measure the individual transit mid-times we fixed all of
the global parameters  
($R_C/R_A$, $a/R_A$, $P$, $i$ and the limb-darkening
coefficients) and fitted each transit event 
separately using the MCMC algorithm described in
\S~\ref{sec:lcfit}. Table~\ref{tab:midtimes} lists the time at the
mid-point 
of each transit $T_{\rm mid}$; the difference between the measured
values and the those predicted by a linear ephemeris; and the formal
measurement uncertainties which range from 10--23 seconds. We see no
statistically significant timing variations.

\subsection{Limits on the Secondary Eclipse Depth}
\label{sec:depth}

To place an upper limit on the secondary eclipse depth we fitted
a simplified transit model to the \kep\ photometry near one-half phase away
from the primary eclipse. We fixed the transit parameters from the fit to
the primary eclipse, along with $e$ and $\omega$ the  from the RV
analysis; assumed no limb darkening; and used final value 
of $\Delta K_P = \dkp \pm \dkpe$. By
allowing only the transit depth, $\sigma_w$ and the terms describing
the out-of-transit 
normalization to vary, we place a
95\% upper limit on the eclipse depth of $d < $~\dup. 

\subsection{Limits on the System Age}

\begin{figure}[!t]
\epsscale{1.1}
\plotone{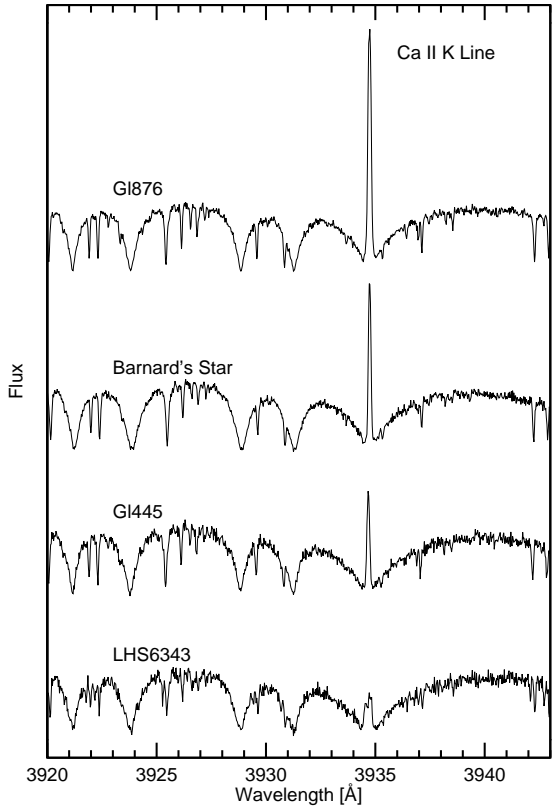}
\caption{HIRES spectra near the CaII K line for \lhs\ and several M
  dwarfs with ages greater than 1 Gyr. \label{fig:cahk}}
\end{figure}

Measuring the ages of M-type stars is notoriously difficult, except for
the rare cases when stars are in clusters or associations, or show
indications of extreme youth. The space motion of \lhs\ rules out membership in any 
known moving group or cluster. We therefore must rely on spectroscopic
and kinematic indicators that can at least tell us if a star is either
younger or older than a few Gyr.

Our HIRES spectra show remarkably low chromospheric activity in the
Ca\,II\,K line (Fig.~\ref{fig:cahk}). None of the 140 low-mass stars 
on the CPS M dwarf survey have a chromospheric S values, as measured
on the Mt. Wilson scale \citep{wright04b}, as low as what we measure 
for \lhs\,A. After attempting to correct for the $\approx30$\%
contamination from star B to the emission core, we estimate $S \approx
0.4$ for \lhs\,A. The most chromospherically quite M dwarf in the
CPS sample is Gl\,445, with an average of $S = 0.5$. 

\lhs\ was not detected by
ROSAT\footnote{The Rontgensatellit (ROSAT) was a joint German, US,
  and British X-ray observatory operational from 1990 to 1999.}, but
at $\d$~pc this only rules out extremely young ages less 
than $\sim100$~Myr. Our HIRES spectra exhibit narrow lines indicating
\vsini~$< 2$~\ks. No rotational modulation of star spots is seen in
the \kep\ light curve, indicating either a long rotation period
($P_{\rm rot} \gtrsim 40$~days) or extremely low spot coverage, or
both. Thus, the age of the \lhs\ system is most likely greater than
1-2 Gyr.  

\begin{figure}[!t]
\epsscale{1.2}
\plotone{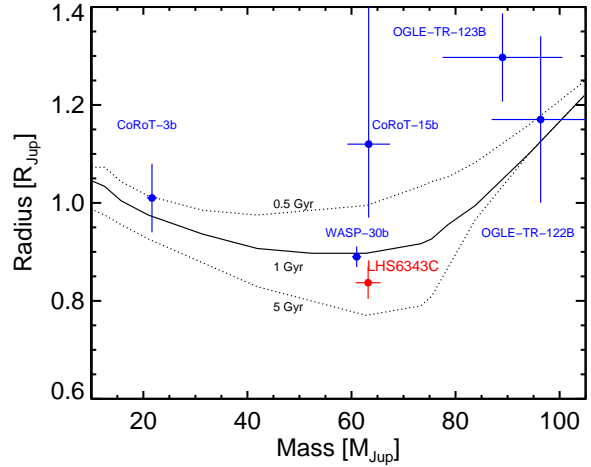}
\caption{The masses and radii of known low-mass, transiting objects.
Also plotted are the predicted mass-radius curves from the
\citet{baraffe98} interior structure models for {0.5, 1, 5}~Gyr, top
to bottom. Not shown are the young, eclipsing brown dwarfs in the
2MASS~2053-05 system \citep{stassun06}, which have radii well above
the plot range. \label{fig:mr}} 
\end{figure}

\section{Discussion}

\lhs,C is remarkably similar to the brown dwarf NLTT41135\,B, recently  
discovered by the {\it MEarth} transit survey \citep{irwin10}. Both
brown dwarfs 
orbit one component of a nearby M+M visual binary, both stellar
systems are hierarchical triples, and both were discovered through
transit photometry. However, owing to
its more favorable orbital inclination, transit photometry provides a
direct measurement of the radius \lhs\,C.

The large mass ratio between parent star and companion argues against
the formation of \lhs\,C in the protoplanetary disk of \lhs\,A. This
is because there is likely not enough material in the disks of M
dwarfs to build a \mc~\mjup\ companion, whether through 
core accretion \citep{laughlin04} or disk instability
\citep{boss06}. It is therefore more likely that \lhs\,C formed in a
similar manner as \lhs\,B, through the fragmentation of a portion of 
their natal 
molecular cloud. \citet{irwin10} make similar arguments regarding the
origin of NLTT\,41135\,B. 

Prior to this year, the only other brown dwarfs
with accurate radius measurements were the substellar components of the  
2MASS\,J05352184-0546085 eclipsing binary system
\citet{stassun06}, and Corot-3\,b \citep{corot3}. The
2MASS\,J05352184-0546085 system was discovered in a young star-forming 
region, and the two brown dwarfs have very large radii for their
masses (0.669~\rsun\ and 0.611~\rsun) because they are still
undergoing gravitational collapse. 

In the latter portion of this year, two other transiting, field brown
dwarfs have been discovered by transit surveys. The ground-based Wide
Angle Survey for Planets discovered WASP-30\,b 
\citep[$60.96 \pm 0.89$~\mjup, $0.89 \pm 0.21$~\rjup;][]{anderson10},
and the space-borne \emph{CoRoT} mission discovered CoRoT-15\,b \citep[$63.3
\pm 1.1$~\mjup, $1.12^{+0.3}_{-0.15}$~\rjup;][]{bouchy10}. Both of these companions
have masses comparable to \lhs\,C, yet orbit single, F-type
stars that are much more massive than \lhs\,A. The current sample of
known transiting brown
dwarfs are shown in Fig.~\ref{fig:mr}, along with two
very low-mass eclipsing M dwarfs. Also shown are the mass and radius
predictions of the \citet{baraffe98} interior models for three
different ages, illustrating that the radius of \lhs\,C is consistent
with the model predictions for a brown dwarf with an age $> 1$~Gyr yet
$< 5$~Gyr. 

Our knowledge of the physical characteristics of brown dwarfs is
starting to expand owing to the growing productivity and efficiency of
wide-field transit surveys, both from the ground and in 
space. \lhs\ is one of the 
closest stars in the \kep\ field and points the way toward
additional brown dwarf discoveries in the near future.  

\acknowledgements
We thank Keivan Stassun and Leslie Hebb for providing us with their 
$B$ and $V$ photometry of the \lhs\ system. We thank Jonathan Irwin
for his independent analysis of the Kepler 
light curve and for encouraging us to incorporate the third-light
correction into our forward-modeling procedure; Josh Carter
for independently confirming our best-fitting light curve parameters;
and John Gizis for pointing out an error in our calculation of
the brown dwarf mass in an earlier draft of this manuscript. 
We gratefully acknowledge the tireless dedication and hard work of the
\kep\ team, without whom this project would not be possible. In
particular, we thank Jon Jenkins and Lucianne Walkowicz for confirming
the planet-like nature of the transits following the initial
identification of transit events by K.A. We gratefully acknowledge the
efforts and dedication 
of the Keck Observatory staff, especially Grant Hill, Scott Dahm and
Hien Tran for their support of 
HIRES and Greg Wirth for support of remote observing. We made use of
the SIMBAD database operated at CDS, 
Strasbourge, France, and NASA's 
Astrophysics Data System Bibliographic Services.
A.\,W.\,H.\ gratefully acknowledges support from a
Townes Post-doctoral Fellowship at the U.\,C.\ Berkeley Space Sciences 
Laboratory. G.\,W.\,M.\ acknowledges NASA grant NNX06AH52G.   
Finally, we extend special thanks to those of
Hawaiian ancestry on whose sacred mountain of Mauna Kea we are
privileged to be guests.   
Without their generous hospitality, the Keck observations presented herein
would not have been possible.

\bibliography{}

\begin{thebibliography}{66}

\bibitem[{{Allard} {et~al.}(2001){Allard}, {Hauschildt}, {Alexander},  {Tamanai}, \& {Schweitzer}}]{allard01}
{Allard}, F., {et~al.} 2001, \apj, 556, 357

\bibitem[{{Anderson} {et~al.}(2010){Anderson}, {Collier Cameron}, {Hellier},  {Lendl}, {Maxted}, {Pollacco}, {Queloz}, {Smalley}, {Smith}, {Todd},  {Triaud}, {West}, {Barros}, {Enoch}, {Gillon}, {Lister}, {Pepe},  {S{\'e}gransan}, {Street}, \& {Udry}}]{anderson10}
{Anderson}, D.~R., {et~al.} 2010, ArXiv e-prints

\bibitem[{{Baraffe} {et~al.}(1998){Baraffe}, {Chabrier}, {Allard}, \&  {Hauschildt}}]{baraffe98}
{Baraffe}, I., {et~al.} 1998, \aap,  337, 403

\bibitem[{{Basri}(2006)}]{basri06}
{Basri}, G. 2006, Astronomische Nachrichten, 327, 3

\bibitem[{{Basri} {et~al.}(1996){Basri}, {Marcy}, \& {Graham}}]{basri96}
{Basri}, G., {Marcy}, G.~W., \& {Graham}, J.~R. 1996, \apj, 458, 600

\bibitem[{{Batalha} {et~al.}(2010){Batalha}, {Borucki}, {Koch}, {Bryson},  {Haas}, {Brown}, {Caldwell}, {Hall}, {Gilliland}, {Latham}, {Meibom}, \&  {Monet}}]{kic}
{Batalha}, N.~M., {et~al.} 2010, \apjl, 713, L109

\bibitem[{{Borucki} {et~al.}(2010){Borucki}, {Koch}, {Basri}, {Batalha},  {Brown}, {Caldwell}, {Caldwell}, {Christensen-Dalsgaard}, {Cochran},  {DeVore}, {Dunham}, {Dupree}, {Gautier}, {Geary}, {Gilliland}, {Gould},  {Howell}, {Jenkins}, {Kondo}, {Latham}, {Marcy}, {Meibom}, {Kjeldsen},  {Lissauer}, {Monet}, {Morrison}, {Sasselov}, {Tarter}, {Boss}, {Brownlee},  {Owen}, {Buzasi}, {Charbonneau}, {Doyle}, {Fortney}, {Ford}, {Holman},  {Seager}, {Steffen}, {Welsh}, {Rowe}, {Anderson}, {Buchhave}, {Ciardi},  {Walkowicz}, {Sherry}, {Horch}, {Isaacson}, {Everett}, {Fischer}, {Torres},  {Johnson}, {Endl}, {MacQueen}, {Bryson}, {Dotson}, {Haas}, {Kolodziejczak},  {Van Cleve}, {Chandrasekaran}, {Twicken}, {Quintana}, {Clarke}, {Allen},  {Li}, {Wu}, {Tenenbaum}, {Verner}, {Bruhweiler}, {Barnes}, \&  {Prsa}}]{kepler10}
{Borucki}, W.~J., {et~al.} 2010, Science, 327, 977

\bibitem[{{Boss}(2006)}]{boss06}
{Boss}, A.~P. 2006, \apj, 643, 501

\bibitem[{{Bouchy} {et~al.}(2010){Bouchy}, {Deleuil}, {Guillot}, {Aigrain},  {Carone}, \& {Cochran}}]{bouchy10}
{Bouchy}, F., {et~al.} 2010, arxiv:1010.0179

\bibitem[{{Bowler} {et~al.}(2009){Bowler}, {Liu}, \& {Cushing}}]{bowler09}
{Bowler}, B.~P., {Liu}, M.~C., \& {Cushing}, M.~C. 2009, \apj, 706, 1114

\bibitem[{{Burgasser} {et~al.}(1999){Burgasser}, {Kirkpatrick}, {Brown},  {Reid}, {Gizis}, {Dahn}, {Monet}, {Beichman}, {Liebert}, {Cutri}, \&  {Skrutskie}}]{burgasser99}
{Burgasser}, A.~J., {et~al.} 1999, \apjl, 522, L65

\bibitem[{{Burgasser} {et~al.}(2007){Burgasser}, {Reid}, {Siegler}, {Close},  {Allen}, {Lowrance}, \& {Gizis}}]{burgasser07}
{Burgasser}, A.~J., {et~al.} 2007, Protostars and Planets V, 427

\bibitem[{{Burrows} {et~al.}(1997){Burrows}, {Marley}, {Hubbard}, {Lunine},  {Guillot}, {Saumon}, {Freedman}, {Sudarsky}, \& {Sharp}}]{burrows97}
{Burrows}, A., {et~al.} 1997, \apj, 491,  856

\bibitem[{{Carter} \& {Winn}(2009)}]{carter09}
{Carter}, J.~A. \& {Winn}, J.~N. 2009, \apj, 704, 51

\bibitem[{{Charbonneau} {et~al.}(2005){Charbonneau}, {Allen}, {Megeath},  {Torres}, {Alonso}, {Brown}, {Gilliland}, {Latham}, {Mandushev}, {O'Donovan},  \& {Sozzetti}}]{charbonneau05}
{Charbonneau}, D., {et~al.} 2005, \apj, 626, 523

\bibitem[{{Claret}(2004)}]{claret04}
{Claret}, A. 2004, \aap, 428, 1001

\bibitem[{{Cushing} \& {Vacca}(2006)}]{cushing06}
{Cushing}, M.~C. \& {Vacca}, W.~D. 2006, \aj, 131, 1797

\bibitem[{{Deleuil} {et~al.}(2008){Deleuil}, {Deeg}, {Alonso}, {Bouchy},  {Rouan}, {Auvergne}, {Baglin}, {Aigrain}, {Almenara}, {Barbieri}, {Barge},  {Bruntt}, {Bord{\'e}}, {Collier Cameron}, {Csizmadia}, {de La Reza},  {Dvorak}, {Erikson}, {Fridlund}, {Gandolfi}, {Gillon}, {Guenther}, {Guillot},  {Hatzes}, {H{\'e}brard}, {Jorda}, {Lammer}, {L{\'e}ger}, {Llebaria},  {Loeillet}, {Mayor}, {Mazeh}, {Moutou}, {Ollivier}, {P{\"a}tzold}, {Pont},  {Queloz}, {Rauer}, {Schneider}, {Shporer}, {Wuchterl}, \& {Zucker}}]{corot3}
{Deleuil}, M., {et~al.} 2008, \aap,  491, 889

\bibitem[{{Delfosse} {et~al.}(2000){Delfosse}, {Forveille}, {S{\'e}gransan},  {Beuzit}, {Udry}, {Perrier}, \& {Mayor}}]{delfosse00}
{Delfosse}, X., {et~al.} 2000, \aap, 364, 217

\bibitem[{{Delorme} {et~al.}(2008){Delorme}, {Willott}, {Forveille},  {Delfosse}, {Reyl{\'e}}, {Bertin}, {Albert}, {Artigau}, {Robin}, {Allard},  {Doyon}, \& {Hill}}]{delorme08}
{Delorme}, P., {et~al.} 2008, \aap, 484, 469

\bibitem[{{Droege} {et~al.}(2006){Droege}, {Richmond}, {Sallman}, \&  {Creager}}]{tass}
{Droege}, T.~F., {et~al.} 2006,  \pasp, 118, 1666

\bibitem[{{Eastman} {et~al.}(2010){Eastman}, {Siverd}, \& {Gaudi}}]{eastman10}
{Eastman}, J., {Siverd}, R., \& {Gaudi}, B.~S. 2010, \pasp, 122, 935

\bibitem[{{Fischer} \& {Valenti}(2005)}]{fischer05b}
{Fischer}, D.~A. \& {Valenti}, J. 2005, \apj, 622, 1102

\bibitem[{{Ford}(2005)}]{ford05}
{Ford}, E.~B. 2005, \aj, 129, 1706

\bibitem[{{Geman} \& {Geman}(1984)}]{geman84}
{Geman}, S. \& {Geman}, D. 1984, IEEE Transactions on Pattern Analysis and  Machine Intelligence, PAMI-6, 721

\bibitem[{{Girardi} {et~al.}(2002){Girardi}, {Bertelli}, {Bressan}, {Chiosi},  {Groenewegen}, {Marigo}, {Salasnich}, \& {Weiss}}]{girardi02}
{Girardi}, L., {et~al.} 2002, \aap, 391, 195

\bibitem[{{Grether} \& {Lineweaver}(2006)}]{grether06}
{Grether}, D. \& {Lineweaver}, C.~H. 2006, \apj, 640, 1051

\bibitem[{{Hayward} {et~al.}(2001){Hayward}, {Brandl}, {Pirger}, {Blacken},  {Gull}, {Schoenwald}, \& {Houck}}]{hayward01}
{Hayward}, T.~L., {et~al.} 2001, \pasp, 113, 105

\bibitem[{{Howard} {et~al.}(2010){Howard}, {Johnson}, {Marcy}, {Fischer},  {Wright}, {Bernat}, {Henry}, {Peek}, {Isaacson}, {Apps}, {Endl}, {Cochran},  {Valenti}, {Anderson}, \& {Piskunov}}]{howard10a}
{Howard}, A.~W., {et~al.} 2010, \apj, 721, 1467

\bibitem[{{Irwin} {et~al.}(2010){Irwin}, {Buchhave}, {Berta}, {Charbonneau},  {Latham}, {Burke}, {Esquerdo}, {Everett}, {Holman}, {Nutzman}, {Berlind},  {Calkins}, {Falco}, {Winn}, {Johnson}, \& {Gazak}}]{irwin10}
{Irwin}, J., {et~al.} 2010, \apj, 718, 1353

\bibitem[{{Jenkins} {et~al.}(2010a){Jenkins}, {Borucki}, {Koch},  {Marcy}, {Cochran}, {Welsh}, {Basri}, {Batalha}, {Buchhave}, {Brown},  {Caldwell}, {Dunham}, {Endl}, {Fischer}, {Gautier}, {Geary}, {Gilliland},  {Howell}, {Isaacson}, {Johnson}, {Latham}, {Lissauer}, {Monet}, {Rowe},  {Sasselov}, {Howard}, {MacQueen}, {Orosz}, {Chandrasekaran}, {Twicken},  {Bryson}, {Quintana}, {Clarke}, {Li}, {Allen}, {Tenenbaum}, {Wu}, {Meibom},  {Klaus}, {Middour}, {Cote}, {McCauliff}, {Girouard}, {Gunter}, {Wohler},  {Hall}, {Ibrahim}, {Kamal Uddin}, {Wu}, {Bhavsar}, {Van Cleve}, {Pletcher},  {Dotson}, \& {Haas}}]{jenkins10c}
{Jenkins}, J.~M., {et~al.} 2010a, \apj, 724, 1108

\bibitem[{{Jenkins} {et~al.}(2010b){Jenkins}, {Caldwell},  {Chandrasekaran}, {Twicken}, {Bryson}, {Quintana}, {Clarke}, {Li}, {Allen},  {Tenenbaum}, {Wu}, {Klaus}, {Middour}, {Cote}, {McCauliff}, {Girouard},  {Gunter}, {Wohler}, {Sommers}, {Hall}, {Uddin}, {Wu}, {Bhavsar}, {Van Cleve},  {Pletcher}, {Dotson}, {Haas}, {Gilliland}, {Koch}, \& {Borucki}}]{jenkins10a}
{Jenkins}, J.~M., {et~al.} 2010b, \apjl, 713, L87

\bibitem[{{Jenkins} {et~al.}(2010c){Jenkins}, {Caldwell},  {Chandrasekaran}, {Twicken}, {Bryson}, {Quintana}, {Clarke}, {Li}, {Allen},  {Tenenbaum}, {Wu}, {Klaus}, {Van Cleve}, {Dotson}, {Haas}, {Gilliland},  {Koch}, \& {Borucki}}]{jenkins10b}
{Jenkins}, J.~M., {et~al.} 2010c, \apjl, 713, L120

\bibitem[{{Johnson}(2009)}]{johnson09rev}
{Johnson}, J.~A. 2009, \pasp, 121, 309

\bibitem[{{Johnson} {et~al.}(2010){Johnson}, {Aller}, {Howard}, \&  {Crepp}}]{johnson10c}
{Johnson}, J.~A., {et~al.} 2010,  \pasp, 122, 905

\bibitem[{{Johnson} \& {Apps}(2009)}]{johnson09b}
{Johnson}, J.~A. \& {Apps}, K. 2009, \apj, 699, 933

\bibitem[{{Johnson} {et~al.}(2004){Johnson}, {Marcy}, {Hamilton}, {Herbst}, \&  {Johns-Krull}}]{johnson04b}
{Johnson}, J.~A., {et~al.} 2004, \aj, 128, 1265

\bibitem[{{Johnson} {et~al.}(2008){Johnson}, {Winn}, {Narita}, {Enya},  {Williams}, {Marcy}, {Sato}, {Ohta}, {Taruya}, {Suto}, {Turner}, {Bakos},  {Butler}, {Vogt}, {Aoki}, {Tamura}, {Yamada}, {Yoshii}, \&  {Hidas}}]{johnson08b}
{Johnson}, J.~A., {et~al.} 2008, \apj, 686, 649

\bibitem[{{Kipping}(2010)}]{kipping10}
{Kipping}, D.~M. 2010, \mnras, 408, 1758

\bibitem[{{Koch} {et~al.}(2010){Koch}, {Borucki}, {Basri}, {Batalha}, {Brown},  {Caldwell}, {Christensen-Dalsgaard}, {Cochran}, {DeVore}, {Dunham},  {Gautier}, {Geary}, {Gilliland}, {Gould}, {Jenkins}, {Kondo}, {Latham},  {Lissauer}, {Marcy}, {Monet}, {Sasselov}, {Boss}, {Brownlee}, {Caldwell},  {Dupree}, {Howell}, {Kjeldsen}, {Meibom}, {Morrison}, {Owen}, {Reitsema},  {Tarter}, {Bryson}, {Dotson}, {Gazis}, {Haas}, {Kolodziejczak}, {Rowe}, {Van  Cleve}, {Allen}, {Chandrasekaran}, {Clarke}, {Li}, {Quintana}, {Tenenbaum},  {Twicken}, \& {Wu}}]{koch10}
{Koch}, D.~G., {et~al.} 2010, \apjl, 713, L79

\bibitem[{{Kratter} {et~al.}(2010){Kratter}, {Murray-Clay}, \&  {Youdin}}]{kratter10}
{Kratter}, K.~M., {Murray-Clay}, R.~A., \& {Youdin}, A.~N. 2010, \apj, 710,  1375

\bibitem[{{Laughlin} {et~al.}(2004){Laughlin}, {Bodenheimer}, \&  {Adams}}]{laughlin04}
{Laughlin}, G., {Bodenheimer}, P., \& {Adams}, F.~C. 2004, \apjl, 612, L73

\bibitem[{{Lawrence} {et~al.}(2007){Lawrence}, {Warren}, {Almaini}, {Edge},  {Hambly}, {Jameson}, {Lucas}, {Casali}, {Adamson}, {Dye}, {Emerson},  {Foucaud}, {Hewett}, {Hirst}, {Hodgkin}, {Irwin}, {Lodieu}, {McMahon},  {Simpson}, {Smail}, {Mortlock}, \& {Folger}}]{ukidss}
{Lawrence}, A., {et~al.} 2007, \mnras, 379, 1599

\bibitem[{{Liu} {et~al.}(2008){Liu}, {Dupuy}, \& {Ireland}}]{liu08b}
{Liu}, M.~C., {Dupuy}, T.~J., \& {Ireland}, M.~J. 2008, \apj, 689, 436

\bibitem[{{Liu} \& {Leggett}(2005)}]{liu05}
{Liu}, M.~C. \& {Leggett}, S.~K. 2005, \apj, 634, 616

\bibitem[{{L{\'o}pez-Morales}(2007)}]{lopez07}
{L{\'o}pez-Morales}, M. 2007, \apj, 660, 732

\bibitem[{{Mandel} \& {Agol}(2002)}]{mandelagol}
{Mandel}, K. \& {Agol}, E. 2002, \apjl, 580, L171

\bibitem[{{Maness} {et~al.}(2007){Maness}, {Marcy}, {Ford}, {Hauschildt},  {Shreve}, {Basri}, {Butler}, \& {Vogt}}]{maness07}
{Maness}, H.~L., {et~al.} 2007, \pasp, 119, 90

\bibitem[{{Marcy} \& {Butler}(2000)}]{marcy00}
{Marcy}, G.~W. \& {Butler}, R.~P. 2000, \pasp, 112, 137

\bibitem[{{Martin} {et~al.}(1997){Martin}, {Basri}, {Delfosse}, \&  {Forveille}}]{martin97}
{Martin}, E.~L., {et~al.} 1997, \aap,  327, L29

\bibitem[{{McCarthy} \& {Zuckerman}(2004)}]{mccarthy04}
{McCarthy}, C. \& {Zuckerman}, B. 2004, \aj, 127, 2871

\bibitem[{{Oppenheimer} {et~al.}(1995){Oppenheimer}, {Kulkarni}, {Matthews}, \&  {Nakajima}}]{oppenheimer95}
{Oppenheimer}, B.~R., {et~al.} 1995, Science, 270, 1478

\bibitem[{{Reid} {et~al.}(2004){Reid}, {Cruz}, {Allen}, {Mungall}, {Kilkenny},  {Liebert}, {Hawley}, {Fraser}, {Covey}, {Lowrance}, {Kirkpatrick}, \&  {Burgasser}}]{reid04}
{Reid}, I.~N., {et~al.} 2004, \aj, 128, 463

\bibitem[{{Ribas}(2006)}]{ribas06}
{Ribas}, I. 2006, \apss, 304, 89

\bibitem[{{Schlaufman} \& {Laughlin}(2010)}]{sl10}
{Schlaufman}, K.~C. \& {Laughlin}, G. 2010, \aap, 519, A105+

\bibitem[{{Seager} \& {Mall{\'e}n-Ornelas}(2003)}]{seager03}
{Seager}, S. \& {Mall{\'e}n-Ornelas}, G. 2003, \apj, 585, 1038

\bibitem[{{Sing}(2010)}]{sing10}
{Sing}, D.~K. 2010, \aap, 510, A21+

\bibitem[{{Sozzetti} {et~al.}(2007){Sozzetti}, {Torres}, {Charbonneau},  {Latham}, {Holman}, {Winn}, {Laird}, \& {O'Donovan}}]{soz07}
{Sozzetti}, A., {et~al.} 2007, \apj, 664,  1190

\bibitem[{{Stassun} {et~al.}(2006){Stassun}, {Mathieu}, \&  {Valenti}}]{stassun06}
{Stassun}, K.~G., {Mathieu}, R.~D., \& {Valenti}, J.~A. 2006, \nat, 440, 311

\bibitem[{{Torres}(2007)}]{torres07}
{Torres}, G. 2007, \apj, 654, 1095

\bibitem[{{Torres} {et~al.}(2008){Torres}, {Winn}, \& {Holman}}]{torres08}
{Torres}, G., {Winn}, J.~N., \& {Holman}, M.~J. 2008, \apj, 677, 1324

\bibitem[{{Troy} {et~al.}(2000){Troy}, {Dekany}, {Brack}, {Oppenheimer},  {Bloemhof}, {Trinh}, {Dekens}, {Shi}, {Hayward}, \& {Brandl}}]{troy00}
{Troy}, M., {et~al.} 2000, 4007, 31

\bibitem[{{Winn}(2008)}]{winn08r}
{Winn}, J.~N. 2008, ASPC, 398, 101

\bibitem[{{Winn} {et~al.}(2009){Winn}, {Holman}, {Henry}, {Torres}, {Fischer},  {Johnson}, {Marcy}, {Shporer}, \& {Mazeh}}]{winn09c}
{Winn}, J.~N., {et~al.} 2009, \apj,  693, 794

\bibitem[{{Wright} \& {Howard}(2009)}]{wrighthoward}
{Wright}, J.~T. \& {Howard}, A.~W. 2009, \apjs, 182, 205

\bibitem[{{Wright} {et~al.}(2004){Wright}, {Marcy}, {Butler}, \&  {Vogt}}]{wright04b}
{Wright}, J.~T., {et~al.} 2004, \apjs,  152, 261

\end{thebibliography}

\clearpage

\begin{deluxetable*}{lccc}
\tablecaption{System Parameters of LHS\,6343 \label{tab:starpars}}
\tablewidth{0pt}
\tablehead{
  \colhead{Parameter} & 
  \colhead{Value}     &
  \colhead{68.3\% Confidence}     &
  \colhead{Comment}   \\
  \colhead{} & 
  \colhead{}     &
  \colhead{Interval}     &
  \colhead{}  
}
\startdata
\emph{Transit Parameters} & & \\
Orbital Period, $P$~[days] & \p & $\pm \pe $ & A \\
Radius Ratio, $(R_C/R_{\star,A})_{\rm corr}$ & \rrc & $\pm \rrce$ & A \\
Transit Depth, $(R_C/R_{\star,A})^2_{\rm corr}$ & \rrcs & $\pm \rrcse$ & A \\
Scaled semimajor axis, $a_R \equiv a/R_\star$  & \ar & $\pm \are$ & A \\
Orbit inclination, $i$~[deg] & \inc & $\pm \ince$ & A \\
Transit impact parameter, $b$ & \imp & $\pm \impe$ & A \\

 & & \\
\emph{Other Orbital Parameters} & & \\
Semimajor axis between Star A and C~[AU] & \arel &$\pm \arele$ & B \\
Eccentricity  & \ecc & $\pm \ecce$ & C \\
Argument of Periastron $\omega$~[degrees] & \om & $\pm \ome$ & C \\
Velocity semiamplitude $K_A$~[\ks] &  \k &$ \pm \ke $ & C \\
Systemic Radial Velocity $\gamma$~[\ks] & \gam & $\pm \game$ & C \\
 & & \\
\emph{Stellar Parameters} & & \\
$M_A$~[$M_\odot$] & \ma & $\pm \mae$ & D \\
$M_B$~[$M_\odot$] & \mb & $\pm \mbe$ & D \\
$R_{\star,A}$~[$R_\odot$] & \ra & $\pm \rae$ & D \\
$\rho_A$~[$\rho_\odot$] & \rhos  & $\pm \rhose$ & A \\
$\log g_A$~[cgs] & \ga & $\pm \gae$ & B \\
\feh & \fe & $\pm \fee$ & D \\
Distance~[pc] & \d & $\pm \de$ & D \\
$V_A$ [mag] & \va & $\pm \vae$ & E \\
$V_B$ [mag] & \vb & $\pm \vbe$ & E \\
$(B-V)_A$ & \bva &  $\pm \bvae$ & F,G \\
$(B-V)_B$ & \bvb &  $\pm \bvbe$ & F,G \\
$\Delta K_P$ [mag]               & \dkp & $\pm \dkpe$ & F \\
$\Delta Z$ [mag]               & \dz & $\pm \dze$ & F,G \\
 & & \\
\emph{Brown Dwarf Parameters} & & \\
$M_C$~[$M_{Jup}$] & \mc  &   $\pm \mce$  & B,C \\
$R_C$~[$R_{Jup}$] & \rc &   $\pm \rce$  & B \\
Mean density, $\rho_C$~[g cm$^{-3}$] & \rhoc & $\pm \rhoce$ & B,C \\
$\log g_C$~[cgs] & \gc  & $\pm \gce$ & A 
\enddata

\tablecomments{Note.---(A) Determined from the parametric fit to the \kep\
  light curve. (B) Based on group A parameters
  supplemented by the photometric stellar mass determination described in
  \S~\ref{sec:starpars}. (C) Based on our analysis of the Keck/HIRES RV
  measurements. (D) Based on our photometric mass and radius
  determinations described in \S~\ref{sec:starpars}. (E)
  Eqn.~\ref{eqn:fe}. (F) Interpolation of Padova 
  model grids. (G) $(B-V)$ colors from the Padova models have been
  corrected by adding  empirically measured offset of $0.21 \pm
  0.07$~mag. Gunn-Z magnitudes were approximated using SDSS
  $z^\prime$-band, with an additive correction of 0.07~mag
  based on the difference between the model $\Delta J$ compared to
  the measured $\Delta J$ in Table~\ref{tab:photpars}.} 

\end{deluxetable*}

\end{document}